%% file: main.tex
\documentclass[sigconf,letterpaper,10pt]{acmart}

\usepackage{url}
\usepackage[utf8]{inputenc}
\usepackage{xcolor}
\usepackage{amsmath}
\usepackage{xspace}
\usepackage{epsfig}
\usepackage{balance}
\usepackage{booktabs}
\usepackage{tabularx}
\usepackage[font=footnotesize]{subcaption}
\usepackage[font=footnotesize]{caption}
\usepackage{mathtools}
\usepackage{soul}
\usepackage{lipsum}
\usepackage{bm}
\usepackage{enumerate}
\usepackage{natbib}
\usepackage{enumitem}

\usepackage[acronyms,nonumberlist,nopostdot,nomain,nogroupskip,acronymlists={hidden}]{glossaries}
\newglossary[algh]{hidden}{acrh}{acnh}{Hidden Acronyms}
\glsdisablehyper

\usepackage{tikz}
\usepackage{pgfplots}
\pgfplotsset{compat=newest}
\pgfplotsset{plot coordinates/math parser=false}
\newlength\fheight
\newlength\fwidth
\usetikzlibrary{plotmarks,patterns,decorations.pathreplacing,backgrounds,calc,arrows,arrows.meta,spy,matrix,scopes}
\usepgfplotslibrary{patchplots,groupplots}
\usepackage{tikzscale}

\input{acronyms.tex}

\usepackage{algorithm}
\usepackage{algorithmicx}
\usepackage{algcompatible}
\usepackage[noend]{algpseudocode}
\usepackage{booktabs}
\usepackage{soul}

\title{Demo: Intelligent Radar Detection in CBRS Band\\ in the Colosseum Wireless Network Emulator}

\author[D. Villa, D. Uvaydov, L. Bonati, P. Johari, J.M. Jornet, T. Melodia]{Davide Villa, Daniel Uvaydov, Leonardo Bonati, Pedram Johari,\\Josep Miquel Jornet, Tommaso Melodia}
\affiliation{%
  \institution{Institute for the Wireless Internet of Things, Northeastern University, Boston, MA, U.S.A.}
  \city{}
  \state{}
  \country{}
  }
\email{{villa.d, uvaydov.d, l.bonati, p.johari, j.jornet, melodia}@northeastern.edu}

\thanks{This work was partially supported by Keysight Technologies and by the U.S. National Science Foundation under grants CNS-1925601 and CNS-2112471. The views and conclusions contained in this document are those of the authors and should not be interpreted as representing the official policies, either expressed or implied, of Keysight Technologies.}

\begin{abstract}

The ever-growing number of wireless communication devices and technologies demands spectrum-sharing techniques. 
Effective coexistence management is crucial to avoid harmful interference, especially with critical systems like nautical and aerial radars in which incumbent radios operate mission-critical communication links.
In this demo, we showcase a framework that leverages Colosseum, the world's largest wireless network emulator with hardware-in-the-loop, as a playground to study commercial radar waveforms coexisting with a cellular network in \acrshort{cbrs} band in complex environments.
We create an ad-hoc high-fidelity spectrum-sharing scenario for this purpose.
We deploy a cellular network to collect \acrshort{iq} samples with the aim of training an ML agent that runs at the base station.
The agent has the goal of detecting incumbent radar transmissions and vacating the cellular bandwidth to avoid interfering with the radar operations.
Our experiment results show an average detection accuracy of 88\%,
with an average detection time of $137$\:ms.

\end{abstract}

\renewcommand\footnotetextcopyrightpermission[1]{}

\begin{document}


\maketitle


\glsresetall
\glsunset{cast}
\glsunset{usrp}
\glsunset{fpga}
\glsunset{uhd}
\glsunset{ue}
\glsunset{bs}
\glsunset{udp}

\section{Introduction}

The evolution of wireless technology has led to increasingly complex wireless systems design.
This requires optimal resource sharing among expanding user sets, prompting researchers to turn to AI as a promising solution for modern system challenges.
AI-driven solutions have enhanced network efficiency and optimized spectrum utilization, outperforming traditional model-based methods in various channel conditions and SNR scenarios.
%
These advancements enable multiple wireless systems to coexist harmoniously, but challenges remain regarding potential interference,
particularly for applications with critical safety communication links.
%
For instance, 5G RANs could interfere with incumbent radar signals in the $3.55-3.7$\:GHz RF band, necessitating thorough research and mitigation strategies to ensure reliable operation for both systems~\cite{caromi2018detection}.
AI-driven spectrum-sharing and interference mitigation algorithms, running often as AI/ML applications in \glspl{ric} like O-RAN's xApps and rApps, have shown promise in optimizing spectrum utilization and mitigating interference~\cite{polese2023understanding, doro2022dapps}.
However, they often require a reliable testing playground and abundant high-quality data.

\begin{figure}[hb]
\setlength\abovecaptionskip{2pt}
    \centering
    \includegraphics[width=\columnwidth]{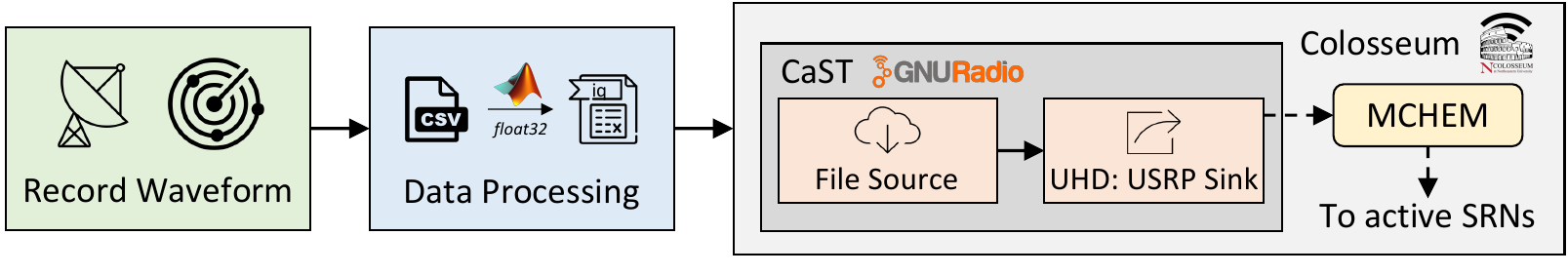}
    \caption{Waveform twinning to integrate signals in Colosseum.}
    \label{fig:waveformblock}
\end{figure}

To address these challenges, we propose a framework for emulating spectrum-sharing scenarios and twinning waveform signals~\cite{villa2023wintech} with both cellular and radar nodes in a high-fidelity digital twin system~\cite{villa2023dt} as depicted in Figure~\ref{fig:waveformblock}.

\noindent The system, implemented on Colosseum~\cite{bonati2021colosseum}, the world's large wireless network emulator with hardware-in-the-loop, provides reliable data collection, AI network training, and realistic scenario testing.
Our framework uses IQ samples of radar and cellular communications to train a \gls{cnn} to detect radar signals and notify the RAN to halt operations, eliminating interference on incumbent radar communications.
Experimental results show an average accuracy of 88\%
with an average detection time of $137$\:ms complying with the minimum detection requirements of commercial transmissions in the CBRS band~\cite{goldreich2016requirements}.

\section{System Design}

We examine the use case of a 4G \gls{ran} transmitting in the \acrshort{cbrs} bandwidth that needs to vacate said bandwidth due to an incumbent radar transmission.
To accomplish this, we design (i) an ad-hoc scenario in Colosseum to collect data and run our experiments, and (ii) a \gls{cnn} for radar detection.

\textbf{Colosseum Waikiki Beach Scenario}
The scenario aims at representing the propagation environment of Waikiki Beach, Honolulu, Hawaii, and has been created through the \acrshort{cast} toolchain~\cite{villa2022cast}.
\begin{figure}[t]
\setlength\abovecaptionskip{1pt}
\setlength\belowcaptionskip{-5pt}
    \centering
    \includegraphics[width=0.78\columnwidth]{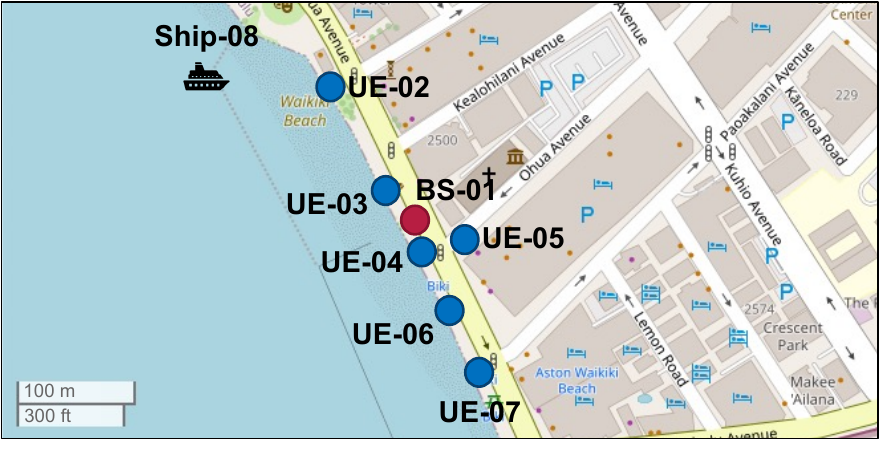}
    \caption{Location of the nodes in the Waikiki Beach scenario.}
    \label{fig:scenarionodes}
\end{figure}
As shown in Figure~\ref{fig:scenarionodes}, the scenario comprises of:
(i) a cellular \gls{bs} (red circle) located at $3$\:m from the ground with GPS coordinates taken from the OpenCellid database of real-world cellular deployments;
(ii) six static \glspl{ue} (blue circles) uniformly distributed in the surroundings of the \gls{bs} at $1$\:m from the ground level emulating hand-held devices;
(iii) one ship (black icon) equipped with a radar, whose antennas are located at a height of $3$\:m and moving in a North-South linear trajectory alongside Waikiki beach at a constant speed of $20$\:knots ($\sim\!\!10$\:m/s).
We leverage the MATLAB ray-tracer to characterize the environment and derive the channel taps among each pair of nodes of our scenario to finally install it in the Colosseum system.

\begin{figure}[ht]
\setlength\abovecaptionskip{1pt}
\setlength\belowcaptionskip{-5pt}
    \centering
    \includegraphics[width=\columnwidth]{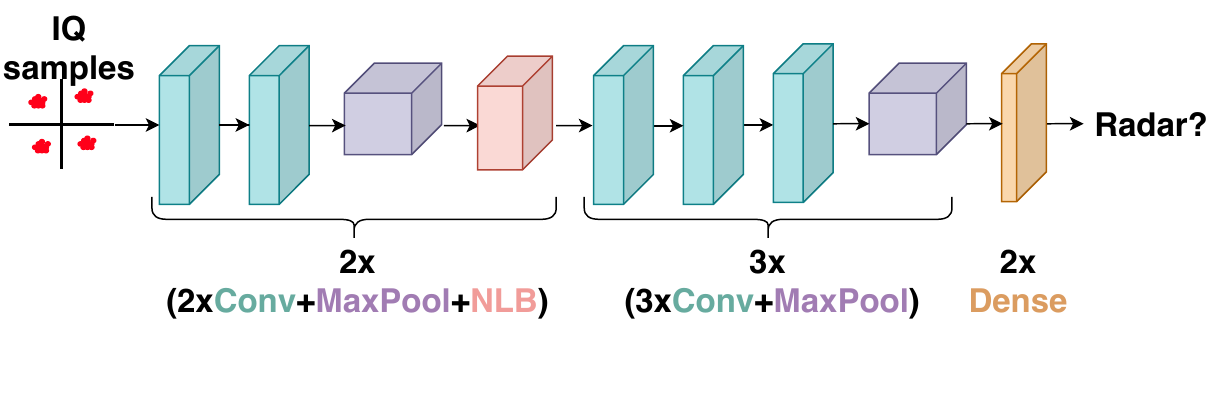}
    \vspace{-0.9cm}
    \caption{\acrshort{cnn} model used to train the radar detector.}
    \label{fig:cnn}
\end{figure}

\textbf{CNN Model}
The \gls{bs} uses an ML inference model trained offline, deployable as a dApp~\cite{doro2022dapps}, to detect radar signals during or before cellular communications.
Figure~\ref{fig:cnn} shows the model architecture of the lightweight \gls{cnn} we leverage which can properly represent our framework's capabilities.

\noindent It consists of a smaller version of VGG16, commonly used in wireless applications, where to the first two convolutional blocks we also append a \gls{nlb}.
The model takes as input \glspl{iq} in the shape of $(batch\_size, 1024, 2)$ and offers a binary label to
each sample: 1 if radar exists, and 0 otherwise.
We collect \glspl{iq} from the Waikiki Beach scenario and we train the ML agent with different combinations and reception gains of radar and cellular signals achieving an average detection accuracy of 88\%.

\section{Demonstration}

\begin{figure}[h]
\setlength\abovecaptionskip{1pt}
\setlength\belowcaptionskip{-5pt}
    \centering
    \includegraphics[width=0.99\columnwidth]{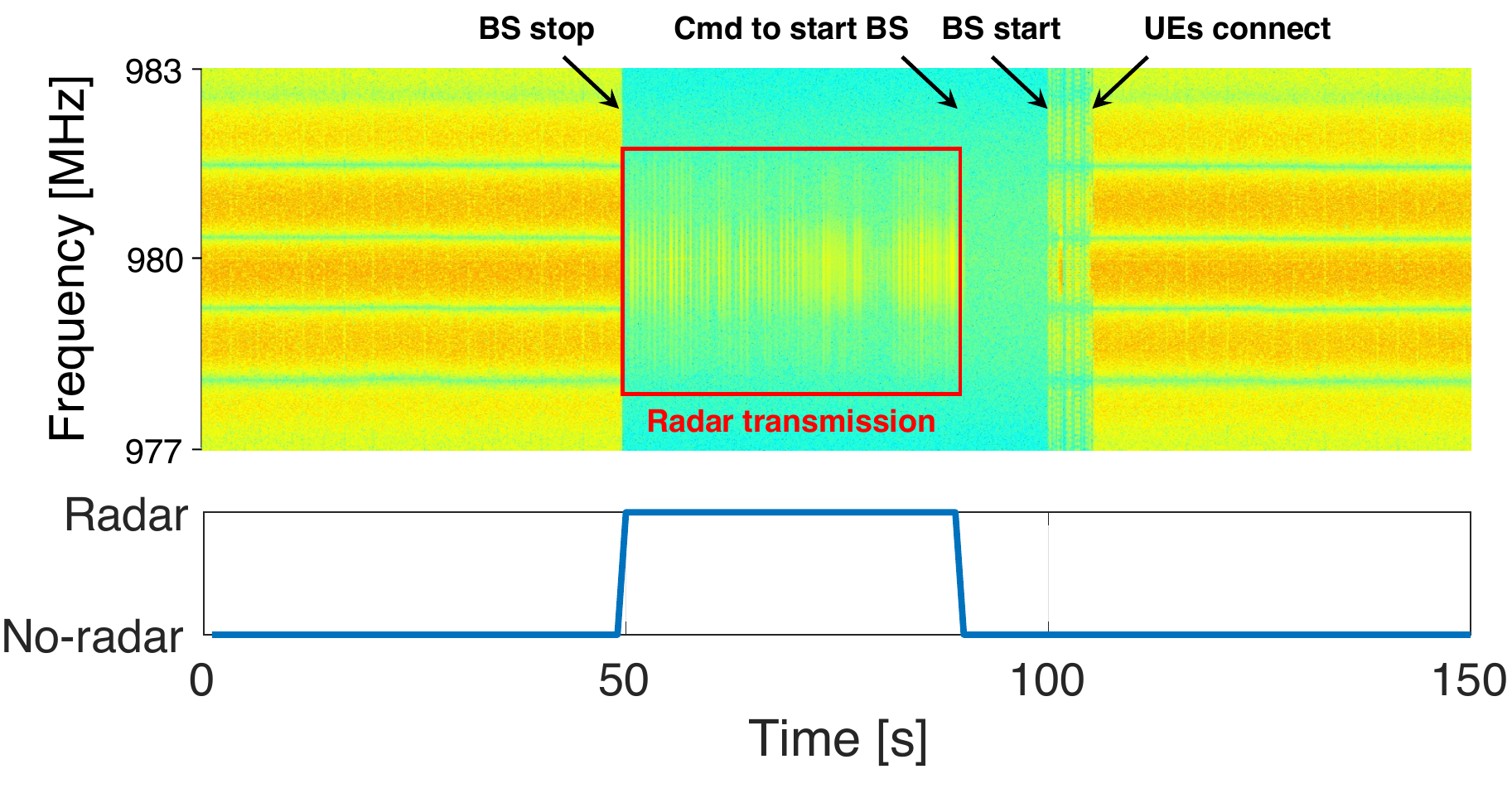}
    \caption{(top) Downlink cellular spectrogram; (bottom) radar detection system result. The \gls{bs} is shut down when a radar transmission is detected and resumes normal operations after no radar is detected.}
    \label{fig:expspectr}
\end{figure}

This demonstration leverages Colosseum and the newly created scenario to deploy a twinned srsRAN-based protocol stack with 1~\gls{bs} and 6~\glspl{ue} and to run a traffic analysis by generating \gls{udp} downlink data streams
through iPerf.
At a certain simulation time, the radar signal is transmitted through a GNU Radio flow graph, designed to support custom radio waveforms.
%
Figure~\ref{fig:expspectr} shows the evolution in time of our demonstration experiment with the top portion displaying the downlink cellular spectrogram centered at $980$\:MHz (i.e., the downlink center frequency that we use for srsRAN in Colosseum), and the bottom one the results of the radar detection system.
From $0$ to $50$\:seconds the \gls{bs} serves \gls{udp} traffic flows to the \glspl{ue}.
At second $50$, the radar starts transmitting and the ML agent detects it and shuts down the \gls{bs}.
When no more radar transmissions are detected (second $90$), the \gls{bs} receives the turn-on command and, after the \glspl{ue} reconnect, it resumes the data streams.
The system leverages a voting system of 100 samples to avoid false positives and false negatives, and a batch size of 10 for a good trade-off between computation time and granularity.
This demonstration proves the effectiveness of our intelligent detector in identifying radar signals and vacating the cellular bandwidth.
%

%
%


\bibliographystyle{ACM-Reference-Format}
\bibliography{biblio}

\end{document}

%% file: acronyms.tex
\newacronym{3gpp}{3GPP}{3rd Generation Partnership Project}
\newacronym{4g}{4G}{4th generation}
\newacronym{5g}{5G}{5th generation}
\newacronym{6g}{6G}{6th generation}
\newacronym{5gc}{5GC}{5G Core}
\newacronym{adc}{ADC}{Analog to Digital Converter}
\newacronym{aerpaw}{AERPAW}{Aerial Experimentation and Research Platform for Advanced Wireless}
\newacronym{ai}{AI}{Artificial Intelligence}
\newacronym{aimd}{AIMD}{Additive Increase Multiplicative Decrease}
\newacronym{am}{AM}{Acknowledged Mode}
\newacronym{amc}{AMC}{Adaptive Modulation and Coding}
\newacronym{amf}{AMF}{Access and Mobility Management Function}
\newacronym{aops}{AOPS}{Adaptive Order Prediction Scheduling}
\newacronym{api}{API}{Application Programming Interface}
\newacronym{apn}{APN}{Access Point Name}
\newacronym{aqm}{AQM}{Active Queue Management}
\newacronym{ausf}{AUSF}{Authentication Server Function}
\newacronym{avc}{AVC}{Advanced Video Coding}
\newacronym{awgn}{AGWN}{Additive White Gaussian Noise}
\newacronym{balia}{BALIA}{Balanced Link Adaptation Algorithm}
\newacronym{bbu}{BBU}{Base Band Unit}
\newacronym{bdp}{BDP}{Bandwidth-Delay Product}
\newacronym{ber}{BER}{Bit Error Rate}
\newacronym{bf}{BF}{Beamforming}
\newacronym{bler}{BLER}{Block Error Rate}
\newacronym{brr}{BRR}{Bayesian Ridge Regressor}
\newacronym{bsr}{BSR}{Buffer Status Report}
\newacronym{bs}{BS}{Base Station}
\newacronym{bpsk}{BPSK}{Binary Phase-shift keying}
\newacronym{bss}{BSS}{Business Support System}
\newacronym{ca}{CA}{Carrier Aggregation}
\newacronym{caas}{CaaS}{Connectivity-as-a-Service}
\newacronym{cb}{CB}{Code Block}
\newacronym{cbrs}{CBRS}{Citizens Broadband Radio Service}
\newacronym{cc}{CC}{Congestion Control}
\newacronym{ccid}{CCID}{Congestion Control ID}
\newacronym{cco}{CC}{Carrier Component}
\newacronym{cdd}{CDD}{Cyclic Delay Diversity}
\newacronym{cdf}{CDF}{Cumulative Distribution Function}
\newacronym{cdn}{CDN}{Content Distribution Network}
\newacronym{cfr}{CFR}{Code of Federal Regulations}
\newacronym{cir}{CIR}{Channel Impulse Response}
\newacronym{cn}{CN}{Core Network}
\newacronym{cnn}{CNN}{Convolutional Neural Network}
\newacronym{codel}{CoDel}{Controlled Delay Management}
\newacronym{comac}{COMAC}{Converged Multi-Access and Core}
\newacronym{cord}{CORD}{Central Office Re-architected as a Datacenter}
\newacronym{cornet}{CORNET}{COgnitive Radio NETwork}
\newacronym{cosmos}{COSMOS}{Cloud Enhanced Open Software Defined Mobile Wireless Testbed for City-Scale Deployment}
\newacronym{cots}{COTS}{Commercial Off-the-Shelf}
\newacronym{cp}{CP}{Control Plane}
\newacronym{cpu}{CPU}{Central Processing Unit}
\newacronym{cqi}{CQI}{Channel Quality Information}
\newacronym{cr}{CR}{Cognitive Radio}
\newacronym{cran}{CRAN}{Cloud \gls{ran}}
\newacronym{crs}{CRS}{Cell Reference Signal}
\newacronym{csi}{CSI}{Channel State Information}
\newacronym{csirs}{CSI-RS}{Channel State Information - Reference Signal}
\newacronym{cu}{CU}{Central Unit}
\newacronym{d2tcp}{D$^2$TCP}{Deadline-aware Data center TCP}
\newacronym{d3}{D$^3$}{Deadline-Driven Delivery}
\newacronym{dac}{DAC}{Digital to Analog Converter}
\newacronym{dag}{DAG}{Directed Acyclic Graph}
\newacronym{darpa}{DARPA}{Defense Advanced Research Projects Agency}
\newacronym{das}{DAS}{Distributed Antenna System}
\newacronym{dash}{DASH}{Dynamic Adaptive Streaming over HTTP}
\newacronym{dc}{DC}{Dual Connectivity}
\newacronym{dccp}{DCCP}{Datagram Congestion Control Protocol}
\newacronym{dce}{DCE}{Direct Code Execution}
\newacronym{dci}{DCI}{Downlink Control Information}
\newacronym{dcl}{DCL}{Dear Colleague Letter}
\newacronym{dctcp}{DCTCP}{Data Center TCP}
\newacronym{dl}{DL}{Downlink}
\newacronym{dmr}{DMR}{Deadline Miss Ratio}
\newacronym{dmrs}{DMRS}{DeModulation Reference Signal}
\newacronym{dod}{DoD}{Department of Defense}
\newacronym{drlcc}{DRL-CC}{Deep Reinforcement Learning Congestion Control}
\newacronym{drs}{DRS}{Discovery Reference Signal}
\newacronym{du}{DU}{Distributed Unit}
\newacronym{e2e}{E2E}{end-to-end}
\newacronym{ecaas}{ECaaS}{Edge-Cloud-as-a-Service}
\newacronym{ecn}{ECN}{Explicit Congestion Notification}
\newacronym{edf}{EDF}{Earliest Deadline First}
\newacronym{em}{EM}{Electromagnetic}
\newacronym{embb}{eMBB}{Enhanced Mobile Broadband}
\newacronym{empower}{EMPOWER}{EMpowering transatlantic PlatfOrms for advanced WirEless Research}
\newacronym{enb}{eNB}{evolved Node Base}
\newacronym{endc}{EN-DC}{E-UTRAN-\gls{nr} \gls{dc}}
\newacronym{epc}{EPC}{Evolved Packet Core}
\newacronym{eps}{EPS}{Evolved Packet System}
\newacronym{es}{ES}{Edge Server}
\newacronym{etsi}{ETSI}{European Telecommunications Standards Institute}
\newacronym[firstplural=Estimated Times of Arrival (ETAs)]{eta}{ETA}{Estimated Time of Arrival}
\newacronym{eutran}{E-UTRAN}{Evolved Universal Terrestrial Access Network}
\newacronym{faas}{FaaS}{Function-as-a-Service}
\newacronym{fapi}{FAPI}{Functional Application Platform Interface}
\newacronym{fcc}{FCC}{Federal Communications Commission}
\newacronym{fdd}{FDD}{Frequency Division Duplexing}
\newacronym{fdm}{FDM}{Frequency Division Multiplexing}
\newacronym{fdma}{FDMA}{Frequency Division Multiple Access}
\newacronym{fed4fire}{FED4FIRE+}{Federation 4 Future Internet Research and Experimentation Plus}
\newacronym{fft}{FFT}{Fast Fourier Transform}
\newacronym{fir}{FIR}{Finite Impulse Response}
\newacronym{fit}{FIT}{Future \acrlong{iot}}
\newacronym{fpga}{FPGA}{Field Programmable Gate Array}
\newacronym{fr2}{FR2}{Frequency Range 2}
\newacronym{fs}{FS}{Fast Switching}
\newacronym{fscc}{FSCC}{Flow Sharing Congestion Control}
\newacronym{ftp}{FTP}{File Transfer Protocol}
\newacronym{fw}{FW}{Flow Window}
\newacronym{ga128}{Ga}{Golay Sequence type A}
\newacronym{ge}{GE}{Gaussian Elimination}
\newacronym{glfsr}{GLFSR}{Galois Linear Feedback Shift Register}
\newacronym{gnb}{gNB}{Next Generation Node Base}
\newacronym{gold}{Gold}{Gold}
\newacronym{gop}{GOP}{Group of Pictures}
\newacronym{gpr}{GPR}{Gaussian Process Regressor}
\newacronym{gpu}{GPU}{Graphics Processing Unit}
\newacronym{gtp}{GTP}{GPRS Tunneling Protocol}
\newacronym{gtpc}{GTP-C}{GPRS Tunnelling Protocol Control Plane}
\newacronym{gtpu}{GTP-U}{GPRS Tunnelling Protocol User Plane}
\newacronym{gtpv2c}{GTPv2-C}{\gls{gtp} v2 - Control}
\newacronym{gw}{GW}{Gateway}
\newacronym{harq}{HARQ}{Hybrid Automatic Repeat reQuest}
\newacronym{hetnet}{HetNet}{Heterogeneous Network}
\newacronym{hh}{HH}{Hard Handover}
\newacronym{hol}{HOL}{Head-of-Line}
\newacronym{hqf}{HQF}{Highest-quality-first}
\newacronym{hss}{HSS}{Home Subscription Server}
\newacronym{http}{HTTP}{HyperText Transfer Protocol}
\newacronym{ia}{IA}{Initial Access}
\newacronym{iab}{IAB}{Integrated Access and Backhaul}
\newacronym{ic}{IC}{Incident Command}
\newacronym{ietf}{IETF}{Internet Engineering Task Force}
\newacronym{ifw}{IFW}{Interference Free Window}
\newacronym{imsi}{IMSI}{International Mobile Subscriber Identity}
\newacronym{imt}{IMT}{International Mobile Telecommunication}
\newacronym{iot}{IoT}{Internet of Things}
\newacronym{ip}{IP}{Internet Protocol}
\newacronym{iq}{IQ}{In-phase and Quadrature}
\newacronym{itu}{ITU}{International Telecommunication Union}
\newacronym{kpi}{KPI}{Key Performance Indicator}
\newacronym{kvm}{KVM}{Kernel-based Virtual Machine}
\newacronym{los}{LOS}{Line-of-Sight}
\newacronym{ls}{LS}{Loosely Synchronised}
\newacronym{lsm}{LSM}{Link-to-System Mapping}
\newacronym{lstm}{LSTM}{Long Short Term Memory}
\newacronym{lte}{LTE}{Long Term Evolution}
\newacronym{lxc}{LXC}{Linux Container}
\newacronym{m2m}{M2M}{Machine to Machine}
\newacronym{mac}{MAC}{Medium Access Control}
\newacronym{manet}{MANET}{Mobile Ad Hoc Network}
\newacronym{mano}{MANO}{Management and Orchestration}
\newacronym{mc}{MC}{Multi-Connectivity}
\newacronym{mcc}{MCC}{Mobile Cloud Computing}
\newacronym{mchem}{MCHEM}{Massive Channel Emulator}
\newacronym{mcs}{MCS}{Modulation and Coding Scheme}
\newacronym{mec}{MEC}{Multi-access Edge Computing}
\newacronym{mec2}{MEC}{Mobile Edge Cloud}
\newacronym{mfc}{MFC}{Mobile Fog Computing}
\newacronym{mi}{MI}{Mutual Information}
\newacronym{mib}{MIB}{Master Information Block}
\newacronym{miesm}{MIESM}{Mutual Information Based Effective SINR}
\newacronym{mimo}{MIMO}{Multiple Input, Multiple Output}
\newacronym{mgen}{MGEN}{Multi-Generator}
\newacronym{ml}{ML}{Machine Learning}
\newacronym{mlr}{MLR}{Maximum-local-rate}
\newacronym[plural=\gls{mme}s,firstplural=Mobility Management Entities (MMEs)]{mme}{MME}{Mobility Management Entity}
\newacronym{mmtc}{mMTC}{Massive Machine-Type Communications}
\newacronym{mmwave}{mmWave}{millimeter wave}
\newacronym{mpdccp}{MP-DCCP}{Multipath Datagram Congestion Control Protocol}
\newacronym{mptcp}{MPTCP}{Multipath TCP}
\newacronym{mr}{MR}{Maximum Rate}
\newacronym{mrdc}{MR-DC}{Multi \gls{rat} \gls{dc}}
\newacronym{mse}{MSE}{Mean Square Error}
\newacronym{mss}{MSS}{Maximum Segment Size}
\newacronym{mt}{MT}{Mobile Termination}
\newacronym{mtd}{MTD}{Machine-Type Device}
\newacronym{mtu}{MTU}{Maximum Transmission Unit}
\newacronym{mumimo}{MU-MIMO}{Multi-user \gls{mimo}}
\newacronym{mvno}{MVNO}{Mobile Virtual Network Operator}
\newacronym{nalu}{NALU}{Network Abstraction Layer Unit}
\newacronym{nas}{NAS}{Network Attached Storage}
\newacronym{nbiot}{NB-IoT}{Narrow Band IoT}
\newacronym{nfv}{NFV}{Network Function Virtualization}
\newacronym{nfvi}{NFVI}{Network Function Virtualization Infrastructure}
\newacronym{nic}{NIC}{Network Interface Card}
\newacronym{nlb}{NLB}{Non-local Block}
\newacronym{nlos}{NLOS}{Non-Line-of-Sight}
\newacronym{now}{NOW}{Non Overlapping Window}
\newacronym{nrdz}{NRDZ}{National Radio Dynamic Zone}
\newacronym{nsf}{NSF}{National Science Foundation}
\newacronym{nsm}{NSM}{Network Service Mesh}
\newacronym[type=hidden]{nr}{NR}{New Radio}
\newacronym{nrf}{NRF}{Network Repository Function}
\newacronym{nsa}{NSA}{Non Stand Alone}
\newacronym{nse}{NSE}{Network Slicing Engine}
\newacronym{nssf}{NSSF}{Network Slice Selection Function}
\newacronym{ntp}{NTP}{Network Time Protocol}
\newacronym{o2i}{O2I}{Outdoor to Indoor}
\newacronym{oai}{OAI}{OpenAirInterface}
\newacronym{oaicn}{OAI-CN}{\gls{oai} \acrlong{cn}}
\newacronym{oairan}{OAI-RAN}{\acrlong{oai} \acrlong{ran}}
\newacronym{oam}{OAM}{Operations, Administration and Maintenance}
\newacronym[plural=\gls{obu}s,firstplural=Onboard Units (OBUs)]{obu}{OBU}{Onboard Unit}
\newacronym{ofdm}{OFDM}{Orthogonal Frequency Division Multiplexing}
\newacronym{olia}{OLIA}{Opportunistic Linked Increase Algorithm}
\newacronym{omec}{OMEC}{Open Mobile Evolved Core}
\newacronym{onap}{ONAP}{Open Network Automation Platform}
\newacronym{onf}{ONF}{Open Networking Foundation}
\newacronym{onos}{ONOS}{Open Networking Operating System}
\newacronym{oom}{OOM}{\gls{onap} Operations Manager}
\newacronym{opnfv}{OPNFV}{Open Platform for \gls{nfv}}
\newacronym[type=hidden]{oran}{O-RAN}{Open \gls{ran}}
\newacronym{orbit}{ORBIT}{Open-Access Research Testbed for Next-Generation Wireless Networks}
\newacronym{os}{OS}{Operating System}
\newacronym{osm}{OSM}{Open Street Map}
\newacronym{oss}{OSS}{Operations Support System}
\newacronym{pa}{PA}{Position-aware}
\newacronym{pase}{PASE}{Prioritization, Arbitration, and Self-adjusting Endpoints}
\newacronym{pawr}{PAWR}{Platforms for Advanced Wireless Research}
\newacronym{pbch}{PBCH}{Physical Broadcast Channel}
\newacronym{pcef}{PCEF}{Policy and Charging Enforcement Function}
\newacronym{pcfich}{PCFICH}{Physical Control Format Indicator Channel}
\newacronym{pcrf}{PCRF}{Policy and Charging Rules Function}
\newacronym{pdcch}{PDCCH}{Physical Downlink Control Channel}
\newacronym{pdcp}{PDCP}{Packet Data Convergence Protocol}
\newacronym{pdsch}{PDSCH}{Physical Downlink Shared Channel}
\newacronym{pdu}{PDU}{Packet Data Unit}
\newacronym{pdp}{PDP}{Power Delay Profile}
\newacronym{pf}{PF}{Proportional Fair}
\newacronym{pgw}{PGW}{Packet Gateway}
\newacronym{phich}{PHICH}{Physical Hybrid ARQ Indicator Channel}
\newacronym{phy}{PHY}{Physical}
\newacronym{pl}{PL}{Path Loss}
\newacronym{pmch}{PMCH}{Physical Multicast Channel}
\newacronym{pmi}{PMI}{Precoding Matrix Indicators}
\newacronym{powder}{POWDER}{Platform for Open Wireless Data-driven Experimental Research}
\newacronym{ppo}{PPO}{Proximal Policy Optimization}
\newacronym{ppp}{PPP}{Poisson Point Process}
\newacronym{prach}{PRACH}{Physical Random Access Channel}
\newacronym{prb}{PRB}{Physical Resource Block}
\newacronym{psd}{PSD}{Power Spectral Density}
\newacronym{psnr}{PSNR}{Peak Signal to Noise Ratio}
\newacronym{pss}{PSS}{Primary Synchronization Signal}
\newacronym{pucch}{PUCCH}{Physical Uplink Control Channel}
\newacronym{pusch}{PUSCH}{Physical Uplink Shared Channel}
\newacronym{qam}{QAM}{Quadrature Amplitude Modulation}
\newacronym{qci}{QCI}{\gls{qos} Class Identifier}
\newacronym{qoe}{QoE}{Quality of Experience}
\newacronym{qos}{QoS}{Quality of Service}
\newacronym{qtgui}{QT-GUI}{QT Graphical User Interface}
\newacronym{quic}{QUIC}{Quick UDP Internet Connections}
\newacronym{rach}{RACH}{Random Access Channel}
\newacronym{ran}{RAN}{Radio Access Network}
\newacronym[firstplural=Radio Access Technologies (RATs)]{rat}{RAT}{Radio Access Technology}
\newacronym{rcn}{RCN}{Research Coordination Network}
\newacronym{rec}{REC}{Radio Edge Cloud}
\newacronym{red}{RED}{Random Early Detection}
\newacronym{renew}{RENEW}{Reconfigurable Eco-system for Next-generation End-to-end Wireless}
\newacronym{rf}{RF}{Radio Frequency}
\newacronym{rfc}{RFC}{Request for Comments}
\newacronym{rfr}{RFR}{Random Forest Regressor}
\newacronym{ric}{RIC}{\gls{ran} Intelligent Controller}
\newacronym{rlc}{RLC}{Radio Link Control}
\newacronym{rlf}{RLF}{Radio Link Failure}
\newacronym{rlnc}{RLNC}{Random Linear Network Coding}
\newacronym{rmse}{RMSE}{Root Mean Squared Error}
\newacronym{rnis}{RNIS}{Radio Network Information Service}
\newacronym{rr}{RR}{Round Robin}
\newacronym{rrc}{RRC}{Radio Resource Control}
\newacronym{rrm}{RRM}{Radio Resource Management}
\newacronym{rru}{RRU}{Remote Radio Unit}
\newacronym{rs}{RS}{Remote Server}
\newacronym{rsrp}{RSRP}{Reference Signal Received Power}
\newacronym{rsrq}{RSRQ}{Reference Signal Received Quality}
\newacronym{rss}{RSS}{Received Signal Strength}
\newacronym{rssi}{RSSI}{Received Signal Strength Indicator}
\newacronym{rsu}{RSU}{Road-Side Unit}
\newacronym{rtt}{RTT}{Round Trip Time}
\newacronym{ru}{RU}{Radio Unit}
\newacronym{rw}{RW}{Receive Window}
\newacronym{rx}{RX}{Receiver}
\newacronym{s1ap}{S1AP}{S1 Application Protocol}
\newacronym{sa}{SA}{standalone}
\newacronym{sack}{SACK}{Selective Acknowledgment}
\newacronym{sap}{SAP}{Service Access Point}
\newacronym{sc2}{SC2}{Spectrum Collaboration Challenge}
\newacronym{scef}{SCEF}{Service Capability Exposure Function}
\newacronym{sch}{SCH}{Secondary Cell Handover}
\newacronym{scoot}{SCOOT}{Split Cycle Offset Optimization Technique}
\newacronym{sctp}{SCTP}{Stream Control Transmission Protocol}
\newacronym{sdap}{SDAP}{Service Data Adaptation Protocol}
\newacronym{sd}{SD}{Standard Deviation}
\newacronym{sdk}{SDK}{Software Development Kit}
\newacronym{sdm}{SDM}{Space Division Multiplexing}
\newacronym{sdma}{SDMA}{Spatial Division Multiple Access}
\newacronym{sdn}{SDN}{Software-defined Networking}
\newacronym{sdr}{SDR}{Software-defined Radio}
\newacronym{seba}{SEBA}{SDN-Enabled Broadband Access}
\newacronym{sgsn}{SGSN}{Serving GPRS Support Node}
\newacronym{sgw}{SGW}{Service Gateway}
\newacronym{si}{SI}{Study Item}
\newacronym{sib}{SIB}{Secondary Information Block}
\newacronym{sinr}{SINR}{Signal to Interference plus Noise Ratio}
\newacronym{sip}{SIP}{Session Initiation Protocol}
\newacronym{siso}{SISO}{Single Input, Single Output}
\newacronym{sla}{SLA}{Service Level Agreement}
\newacronym{sm}{SM}{Saturation Mode}
\newacronym{smf}{SMF}{Session Management Function}
\newacronym{smo}{SMO}{Service Management and Orchestration}
\newacronym{sms}{SMS}{Short Message Service}
\newacronym{smsgmsc}{SMS-GMSC}{\gls{sms}-Gateway}
\newacronym{snr}{SNR}{Signal-to-Noise-Ratio}
\newacronym{son}{SON}{Self-Organizing Network}
\newacronym{sptcp}{SPTCP}{Single Path TCP}
\newacronym{srb}{SRB}{Service Radio Bearer}
\newacronym{srn}{SRN}{Standard Radio Node}
\newacronym{srs}{SRS}{Sounding Reference Signal}
\newacronym{ss}{SS}{Synchronization Signal}
\newacronym{sss}{SSS}{Secondary Synchronization Signal}
\newacronym{st}{ST}{Spanning Tree}
\newacronym{svc}{SVC}{Scalable Video Coding}
\newacronym{tb}{TB}{Transport Block}
\newacronym{tcp}{TCP}{Transmission Control Protocol}
\newacronym{tdd}{TDD}{Time Division Duplexing}
\newacronym{tdm}{TDM}{Time Division Multiplexing}
\newacronym{tdma}{TDMA}{Time Division Multiple Access}
\newacronym{tfl}{TfL}{Transport for London}
\newacronym{tfrc}{TFRC}{TCP-Friendly Rate Control}
\newacronym{tft}{TFT}{Traffic Flow Template}
\newacronym{tgen}{TGEN}{Traffic Generator}
\newacronym{tip}{TIP}{Telecom Infra Project}
\newacronym{tm}{TM}{Transparent Mode}
\newacronym{to}{TO}{Telco Operator}
\newacronym{toa}{ToA}{Time of Arrival}
\newacronym{tr}{TR}{Technical Report}
\newacronym{trp}{TRP}{Transmitter Receiver Pair}
\newacronym{ts}{TS}{Technical Specification}
\newacronym{tti}{TTI}{Transmission Time Interval}
\newacronym{ttt}{TTT}{Time-to-Trigger}
\newacronym{tx}{TX}{Transmitter}
\newacronym{uas}{UAS}{Unmanned Aerial System}
\newacronym{uav}{UAV}{Unmanned Aerial Vehicle}
\newacronym{udm}{UDM}{Unified Data Management}
\newacronym{udp}{UDP}{User Datagram Protocol}
\newacronym{udr}{UDR}{Unified Data Repository}
\newacronym{ue}{UE}{User Equipment}
\newacronym{uhd}{UHD}{\gls{usrp} Hardware Driver}
\newacronym{ul}{UL}{Uplink}
\newacronym{um}{UM}{Unacknowledged Mode}
\newacronym{uml}{UML}{Unified Modeling Language}
\newacronym{upa}{UPA}{Uniform Planar Array}
\newacronym{upf}{UPF}{User Plane Function}
\newacronym{urllc}{URLLC}{Ultra Reliable and Low Latency Communication}
\newacronym{usa}{U.S.}{United States}
\newacronym{usim}{USIM}{Universal Subscriber Identity Module}
\newacronym{usrp}{USRP}{Universal Software Radio Peripheral}
\newacronym{utc}{UTC}{Urban Traffic Control}
\newacronym{vim}{VIM}{Virtualization Infrastructure Manager}
\newacronym{vm}{VM}{Virtual Machine}
\newacronym{vnf}{VNF}{Virtual Network Function}
\newacronym{volte}{VoLTE}{Voice over \gls{lte}}
\newacronym{voltha}{VOLTHA}{Virtual OLT HArdware Abstraction}
\newacronym{vr}{VR}{Virtual Reality}
\newacronym{vran}{vRAN}{Virtualized \gls{ran}}
\newacronym{vss}{VSS}{Video Streaming Server}
\newacronym{wbf}{WBF}{Wired Bias Function}
\newacronym{wf}{WF}{Wired-first}
\newacronym{wi}{WI}{Wireless InSite}
\newacronym{wlan}{WLAN}{Wireless Local Area Network}
\newacronym{wsr}{WSR}{Weather Surveillance Radar}
\newacronym{pnf}{PNF}{Physical Network Function}
\newacronym{drl}{DRL}{Deep Reinforcement Learning}
\newacronym{mtc}{MTC}{Machine-type Communications}
\newacronym{v2x}{V2X}{Vehicle-to-everything}
\newacronym{cast}{CaST}{Channel emulation generator and Sounder Toolchain}

%% file: main.bbl

\begin{thebibliography}{8}


\ifx \showCODEN    \undefined \def \showCODEN     #1{\unskip}     \fi
\ifx \showDOI      \undefined \def \showDOI       #1{#1}\fi
\ifx \showISBNx    \undefined \def \showISBNx     #1{\unskip}     \fi
\ifx \showISBNxiii \undefined \def \showISBNxiii  #1{\unskip}     \fi
\ifx \showISSN     \undefined \def \showISSN      #1{\unskip}     \fi
\ifx \showLCCN     \undefined \def \showLCCN      #1{\unskip}     \fi
\ifx \shownote     \undefined \def \shownote      #1{#1}          \fi
\ifx \showarticletitle \undefined \def \showarticletitle #1{#1}   \fi
\ifx \showURL      \undefined \def \showURL       {\relax}        \fi
\providecommand\bibfield[2]{#2}
\providecommand\bibinfo[2]{#2}
\providecommand\natexlab[1]{#1}
\providecommand\showeprint[2][]{arXiv:#2}

\bibitem[Bonati et~al\mbox{.}(2021)]%
        {bonati2021colosseum}
\bibfield{author}{\bibinfo{person}{L. Bonati}, \bibinfo{person}{P. Johari},
  \bibinfo{person}{M. Polese}, \bibinfo{person}{S. {D'Oro}},
  \bibinfo{person}{S. Mohanti}, \bibinfo{person}{M. Tehrani-Moayyed},
  \bibinfo{person}{D. Villa}, \bibinfo{person}{S. Shrivastava},
  \bibinfo{person}{C. Tassie}, \bibinfo{person}{K. Yoder}, \bibinfo{person}{A.
  Bagga}, \bibinfo{person}{P. Patel}, \bibinfo{person}{V. Petkov},
  \bibinfo{person}{M. Seltzer}, \bibinfo{person}{F. Restuccia},
  \bibinfo{person}{M. Gosain}, \bibinfo{person}{K.~R. Chowdhury},
  \bibinfo{person}{S. Basagni}, {and} \bibinfo{person}{T. Melodia}.}
  \bibinfo{year}{2021}\natexlab{}.
\newblock \showarticletitle{{Colosseum: Large-Scale Wireless Experimentation
  Through Hardware-in-the-Loop Network Emulation}}. In
  \bibinfo{booktitle}{\emph{Proc. of IEEE DySPAN}}.
\newblock


\bibitem[Caromi et~al\mbox{.}(2018)]%
        {caromi2018detection}
\bibfield{author}{\bibinfo{person}{R. Caromi}, \bibinfo{person}{M. Souryal},
  {and} \bibinfo{person}{WB Yang}.} \bibinfo{year}{2018}\natexlab{}.
\newblock \showarticletitle{Detection of Incumbent Radar in the 3.5 GHz CBRS
  Band}. In \bibinfo{booktitle}{\emph{Proc. of IEEE GlobalSIP}}. IEEE.
\newblock


\bibitem[D'Oro et~al\mbox{.}(2022)]%
        {doro2022dapps}
\bibfield{author}{\bibinfo{person}{S. D'Oro}, \bibinfo{person}{M. Polese},
  \bibinfo{person}{L. Bonati}, \bibinfo{person}{H. Cheng}, {and}
  \bibinfo{person}{T. Melodia}.} \bibinfo{year}{2022}\natexlab{}.
\newblock \showarticletitle{{dApps: Distributed Applications for Real-Time
  Inference and Control in O-RAN}}.
\newblock \bibinfo{journal}{\emph{IEEE Communications Magazine}}
  \bibinfo{volume}{60}, \bibinfo{number}{11} (\bibinfo{date}{November}
  \bibinfo{year}{2022}), \bibinfo{pages}{52--58}.
\newblock


\bibitem[Goldreich(2016)]%
        {goldreich2016requirements}
\bibfield{author}{\bibinfo{person}{O. Goldreich}.}
  \bibinfo{year}{2016}\natexlab{}.
\newblock \showarticletitle{{Requirements for Commercial Operation in the U.S.
  3550-3700 MHz Citizens Broadband Radio Service Band}}. In
  \bibinfo{booktitle}{\emph{Wireless Innovation Forum}}.
\newblock


\bibitem[Polese et~al\mbox{.}(2023)]%
        {polese2023understanding}
\bibfield{author}{\bibinfo{person}{M. Polese}, \bibinfo{person}{L. Bonati},
  \bibinfo{person}{S. D'Oro}, \bibinfo{person}{S. Basagni}, {and}
  \bibinfo{person}{T. Melodia}.} \bibinfo{year}{2023}\natexlab{}.
\newblock \showarticletitle{{Understanding O-RAN: Architecture, Interfaces,
  Algorithms, Security, and Research Challenges}}.
\newblock \bibinfo{journal}{\emph{IEEE Communications Surveys \& Tutorials}}
  \bibinfo{volume}{25}, \bibinfo{number}{2} (\bibinfo{date}{January}
  \bibinfo{year}{2023}), \bibinfo{pages}{1376--1411}.
\newblock


\bibitem[Villa et~al\mbox{.}(2022)]%
        {villa2022cast}
\bibfield{author}{\bibinfo{person}{D. Villa}, \bibinfo{person}{M.
  Tehrani-Moayyed}, \bibinfo{person}{P. Johari}, \bibinfo{person}{S. Basagni},
  {and} \bibinfo{person}{T. Melodia}.} \bibinfo{year}{2022}\natexlab{}.
\newblock \showarticletitle{{{CaST}: A Toolchain for Creating and
  Characterizing Realistic Wireless Network Emulation Scenarios}}. In
  \bibinfo{booktitle}{\emph{Proc. of ACM WiNTECH}}. \bibinfo{address}{Sydney,
  Australia}.
\newblock


\bibitem[Villa et~al\mbox{.}(2023a)]%
        {villa2023dt}
\bibfield{author}{\bibinfo{person}{D. Villa}, \bibinfo{person}{M.
  Tehrani-Moayyed}, \bibinfo{person}{C.~P. Robinson}, \bibinfo{person}{L.
  Bonati}, \bibinfo{person}{P. Johari}, \bibinfo{person}{M. Polese},
  \bibinfo{person}{S. Basagni}, {and} \bibinfo{person}{T. Melodia}.}
  \bibinfo{year}{2023}\natexlab{a}.
\newblock \showarticletitle{{Colosseum as a Digital Twin: Bridging Real-World
  Experimentation and Wireless Network Emulation}}.
\newblock \bibinfo{journal}{\emph{arXiv:2303.17063 [cs.NI]}}
  (\bibinfo{date}{March} \bibinfo{year}{2023}), \bibinfo{pages}{1--15}.
\newblock


\bibitem[Villa et~al\mbox{.}(2023b)]%
        {villa2023wintech}
\bibfield{author}{\bibinfo{person}{D. Villa}, \bibinfo{person}{D. Uvaydov},
  \bibinfo{person}{L. Bonati}, \bibinfo{person}{P. Johari},
  \bibinfo{person}{J.~M. Jornet}, {and} \bibinfo{person}{T. Melodia}.}
  \bibinfo{year}{2023}\natexlab{b}.
\newblock \showarticletitle{{Twinning Commercial Radio Waveforms in the
  Colosseum Wireless Network Emulator}}. In \bibinfo{booktitle}{\emph{Proc. of
  ACM WiNTECH}}. \bibinfo{address}{Madrid, Spain}.
\newblock


\end{thebibliography}
